# Estimate of drainage water behaviour in shallow lakes

Léa J. El-Jaick* and A. A. Gomes

Centro Brasileiro de Pesquisas Físicas

Rua Xavier Sigaud 150, CEP 22290-180, Rio de Janeiro, RJ, Brazil.

______________________________________________________________________

**Abstract**
A theoretical estimate of the *explicit* time dependence of a drainage water of shallow lakes is presented as an important contribution for understanding the lake dynamics. This information can be obtained from a sum of functions, largely used in fitting of experimental data. These functions were chosen because their centre and weight yield a good description of the water basin behaviour. The coefficients of these functions are here extracted using results of measured and / or calculated data for the state variables describing the shallow West Lake, Hangzou. This procedure can also be applied to other shallow lakes, generating geological information about their drainage basin, which is one of the most important parameters to describe their micrometeorological behaviour. One concludes this work emphasizing the relevance of the *explicit* time dependence of the drainage variables and the requirement of measured data to validate this approach.

______________________________________________________________________

*Corresponding author: Tel.: +55-21-2141-7178/7154; Fax: +55-21-2141-7540
E-mail address: leajj@cbpf.br

# 1. Introduction

Different approaches for experiments and modelling have been proposed to study the dynamics of lakes in specific environments. The main idea of the present modelling is a more detailed description of lakes, which takes into account the influence of their environment and its corresponding micrometeorological behaviour. For better understanding, one firstly introduces a summary of experimental / theoretical modelling efforts describing details of lakes, in several regions of the world.

One starts with the work about the restoration of the Lake Eymir, in Turkey (Beklioglu et al., 2003) because some aspects discussed are interesting for future studies about lakes of Rio de Janeiro State, Brazil. In effect, these authors show that the nutrient loading is a serious threat to water quality. This successful restoration was executed reducing the areal loading of total phosphorus and dissolved inorganic nitrogen. In spite of this effort, the poor water clarity and mainly low-submerged plant coverage persisted. They also found that some kinds of fish perpetuate the poor water condition. However, a decrease in the inorganic suspended solid, and in the chlorophyll-a concentrations, reduced the fish stock and led to a better water quality. The existence of fishes is not considered in the present work, neither for the lakes of the Rio de Janeiro State (Southeast of Brazil).

Another subject of interest here is the study of the sensibility to acidic deposition in lakes of Adirondack Mountain region of New York State (Civerolo et al., 2003). The ambient air and the lake water quality were analysed over a period of ten years, in order to find the role of seasonality and time variation, in estimating temporal trends in the experimental data.

In addition, a very interesting work concerns the zooplankton community structure (Cottenie et al., 2003). Using data obtained during three years for a system of highly interconnected ponds, these authors studied the influences of regional interactions over the local zooplankton communities, showing that environmental constraints are strongly related with the zooplankton community structure.

Within this general approach, one will consider some examples, which apply different techniques to study the lake environment behaviour.

Different experimental techniques have been used to study the role of fish contamination on lakes and marine waters by chemical products or natural toxins. Liquid chromatography with mass spectrometric detection was developed by Dahlman et al. (2003) to determine various algal and cyanobacterial toxins extracted from phytoplankton, which can lead on shellfish poisoning. Page and Murphy (2003) also used a geographic information system (GIS) approach to create a database to establish the mercury (Hg) levels in many fishes from remote lakes in Canada, where the Hg quantity exceeds the recommended level for human consumption.

This summary gives a panoramic view concerning some shallow lake studies that have been appeared in ecological literature. Another point of view is taken in a paper by Hongping and Jianyi (2002) who describe the ecosystem of West Lake, Hangzhou, using an algal dynamic model that includes several state variables, which are not present in the aforementioned works.

Based on the last paper, the present work uses a more general equations for the water column, including explicitly the time dependence of the drainage basin function $Q(t)$, concerning micrometeorological effects. This generalization clarifies the importance of the role of environment in the lake dynamics and allows obtaining the simulated results found by Hongping and Jianyi (2002).

One presents two different steps: the first one is mostly theoretical description of lake modelling (section 2), and the second concerns the interpretation of experimental data as measured for the lake phyto-zooplankton community, observed on an annual time scale (section 3). As it will be made more clear later on, the calculated data by the authors mentioned above, for the lake community, as a function of time, are the starting point to extract information on the strength and time dependence of the drainage water.

## 2. The model

The dynamical state equations for the description of a lake in *a given* geographical environment involve several parameters. The most frequently used are the rate equations which describe quantitatively the time dependence of the process of birth, growth and decomposition of the lake constituents, together with forcing functions like, for example, the time dependent sunlight incidence. Many of these *rate constants* can be measured using physico-chemical techniques, and the values, which illustrate this study, are those assumed for the Chinese West Lake by Hongping and Jianyi (2002) at their Table 2. Examples suggesting adequate approaches to hydrobiology as applied to lakes, considered here, are given by Ahlgren et al. (1988) and Krivtsov et al. (1999). Obviously, for eventual application to Brazilian lakes, the physico-chemical rate processes should be adapted to the local conditions and again extracted from experiment (Carmouze, 1984; Suzuki, 1997). In particular, shallow lakes connected to the ocean do exist in the region of the North of Rio de Janeiro State. For application to these particular lakes, the dynamic equations adopted in the present work should be extended to include specific effects of the connection with the sea. These cases may be considered in future work.

The structure of the physico-chemical aspects of the model equations is currently assumed *independent* of the geographic coordinates of a given lake. However, for very specific situations, e.g., when it exists a *connection with the sea,* the mathematical structure of the model must be modified. Thus, in some situations exhibited in Chinese West Lake, it is necessary to introduce *explicit time dependence* in some parameters associated with the geographic environment, and from these parameters, the time dependence of the drainage water is extracted using the coefficients of a Lorentzian expansion, as it will be seen below.

One adopts here the assumptions of Hongping and Jianyi (2002) for the modelling of lakes. The *state variables* are the biomasses of four species of algae, Cyanophyta, Chlorophyta, Criptophyta and Bacillariophyta, *BA1(t), BA2(t), BA3(t) and BA4(t)*, respectively, with their respective content of phosphorus *PA1(t), PA2(t), PA3(t) and PA4(t)*; biomass of the zooplankton, *BZ(t)*, and its content of phosphorus, *PZ(t)*; phosphorus in detritus, *PD(t)*; phosphorus in sediment, *PE(t)*; and finally orthophosphate, *PS(t)*, in the lake water column. In the present model, as it was mentioned before, the existence of fishes is entirely disregarded. This assumption may also be applied to lakes of the Southeast of Brazil, but not when studying Brazilian lakes in the North of the Country, where fishery is quite relevant.

The *central point* in this work concerns the distribution in *time* of the drainage water which is considered here as the main micrometeorological parameter. Drainage water *depends on time*, via a distribution function *Q(t)* and this quantity is described by a sum of Lorentzians functions, contrary to Hongping and Jianyi (2002) who calculate the quantity of drawing water *Q*, according to the average quantity every month, which, in fact, means that *Q* is as a *time independent parameter* in their description of the Chinese Lake. The motivation to fit the *Q(t)* functions with a sum of Lorentzians was inspired by the usual approach adopted in techniques of experimental physics, like Electron Paramagnetic Resonance (EPR) and Mössbauer effect. Experimental results are very sensitive to the neighbourhood of the probe. This fact is expressed by a combination of given functions, like the Lorentzians, which are defined by their centres describing, in the present case, the micrometeorological time of rain precipitation, and their widths, which describe the time interval of precipitation and the soil diffusion process time.

The theoretical results obtained by Hongping and Jianyi (2002) were retrieved in order to extract the time dependence of the water drainage. It should be stressed that for other available data corresponding to lakes, in Brazil or else, this proposal can be applied to extract also drainage water results. From now on, the dynamics defined below will be considered fixed and well defined; thus, only parameters and forcing functions may be changed to describe different lakes and regions.

The state equations, from which the information about the drainage water is obtained, have the following *general form*, where the time dependence has been previously defined:

$$\frac{dBA_i}{dt} = \left\{ A_1^{(i)}[t; PS, BA_i] - Q_{BA_i}^{(i)}(t)/V \right\} \times BA_i - BZ \times A_2^{(i)}[t; BA_i] \qquad (1)$$

where $i = 1,2,3,4$ for the four algae and *V* is the volume lake. The main difference in respect to Hongping and Jianyi (2002) concerns the time dependent drainage water as described by $Q_{BA_i}^{(i)}(t)$. These distribution functions are *strictly time dependent* and this intends to describe the contribution of the diverse micrometeorological effects.

For the zooplankton, one has the dynamics:

$$\frac{dBZ}{dt} = B_1[t; BA_i] \times BZ - (Q_{BZ}(t)/V) \times BZ \qquad (2)$$

For orthophosphate and phosphorus in detritus one has respectively

$$\frac{dPS}{dt} = LPS + B_2[t; PD, PS, PE, PA_i, BA_i] - (Q_{PS}(t)/V) \times PS \qquad (3)$$

$$\frac{dPD}{dt} = LPD + B_3(t; PZ, PD) - (Q_{PD}(t)/V) \times PD \qquad (4)$$

It should be noted that the *time dependent* drainage water contributions to the lake dynamics are expected to be different for each one of the lake state variables and represented as *Q(t) / V*. This difference is associated with the several substances drained, which interfere in the time dependence of the state variables.

In these equations the functions, $A_1^{(i)}(t)$, $A_2^{(i)}(t)$, $B_1^{(i)}(t)$, $B_2^{(i)}(t)$, and $B_3(t)$ are strictly non linear in the state variables and time. These functions include also the model parameters that are assumed well defined experimentally. In Eq. (3), $PA_i(t)$ correspond to the phosphorus in the algae. These non-linear functions have the general form $F(t; S_1(t), S_2(t), ... S_q(t); \mathbf{1}_1, \mathbf{1}_2, ... \mathbf{1}_p)$ where $\mathbf{1}_p$ are the parameters of the model and $S_q(t)$ are the state variables obtained from experiment as a function of time  *t*. The parameters $1_p$ are identical to those of Hongping and Jianyi (2002) but a simulation changing them can be made, since the time dependence of the state variable for this new situation is known and thus, using the proposed procedure, the new drainage results can be obtained. The detailed formulation of the functions $A_1^{(i)}(t)$, $A_2^{(i)}(t)$, $B_1^{(i)}(t)$, $B_2^{(i)}(t)$, and $B_3(t)$ are presented in Hongping and Jianyi (2002), with the parameters defined in their Table 2, which one expects to be adequate to the present study. Given the simulated values for time in the interval from *t* = 0 and *t* =360 days, for these functions, and performing numerical differentiation of the available time dependence of the state variables, one can extract the corresponding drainage water *Q(t)´s* that will be fitted by eight Lorentzians. The numerical results of such a procedure is shown in the figures.

Another type of dynamical equation concerns the amount of phosphorus in the four species of algae and the zooplankton. Their equations of motion are written in general terms:

$$\frac{dPA_i}{dt} = C_i[t; BA_i, PS, PA_i] - \left(Q_{BA_i}^{(i)}(t)/V\right) \times PA_i \qquad (5)$$

where *i* = 1,2,3,4

$$\frac{dPZ}{dt} = D[t; BA_i, PA_i, BZ] - \left(Q_{BZ}(t)/V\right) \times PZ \qquad (6)$$

where, again, different $Q_i(t)$ are used since different substances are introduced by the geological drainage distribution.

Since one has no data concerning the internal phosphorus, the values for *PA_i(t)* and *PZ(t)* are assumed proportional to the algae and zooplankton biomasses respectively. This is based on so-called Redfield molecule (Redfield et al., 1963), which intends to represent the average content of this element in these biomasses.

For the case of phosphorus in the sediment, one has the following equation:

$$\frac{dPE}{dt} = E[t; PE, PS, PD, PA_i] \qquad (7)$$

It is important to note that contrary to the remaining equations for the lake state variables, the drainage water *does not appear explicitly* in Eq. (7). This fact seems to be reasonable since drainage, only indirectly affects the sediment of the lake via the water column dynamics, and this is expected to have a *distinct time scale*.

## 3. Numerical results and conclusions

To illustrate the results of the model calculation for the drainage water, which is *time dependent* as suggested by micrometeorology, contrary to Hongping and Jianyi (2002), one decided to fit its *time dependence* through a sum of Lorentzians. As indicated in Fig. 1 of this reference, the water drainage basin has a rather complex geometrical topology, including several distinct water distributions. One emphasizes that this fact was the motivation to adopt this procedure, which is based on the physical experiment fitting of results in complex lattices in solids, studied by EPR or Mössbauer techniques. Such procedure describes the complex structure of the solid by the contributions of a superposition of several Lorentzians. Their centres and widths represent adequately the relative relevance of the involved water paths of Fig. 1 de Hongping and Jianyi (2002). A possible test of this approach is to compare with data to be obtained in Brazilian lakes like those of Campos dos Goytacazes region of the north of Rio de Janeiro State. Eventual description including underground water dynamics may also be a subject of a further work.

The numerical procedure goes as follows: the data presented in the figures of Hongping and Jianyi (2002) for $BA_i(t)$, $i = 1,2,3,4$, $BZ(t)$, $PS(t)$ and $PD(t)$, are taken together with the parameters presented in their Table 2. Given these data, one numerically differentiates the curves adopting the values for the quantities *LPS* and *LPD* given by these authors. These parameters are other forms of phosphorus, which are extracted from *outside* the lake. Thus, it is possible to extract the curves of $Q_{BA_i}^{(i)}(t)$ and $Q_M(t)$ (with $M=BZ$, $PS$ and $PD$) using equations Eq. (1) to Eq. (4). Then, the obtained functions are fitted using the sum of eight Lorentzians of the general form:

$$H(t) = \sum_{i=1,8} Y^i(t) \qquad (8)$$

with

$$Y^{(i)} = Y_0^{(i)} + \frac{2A^i}{p} \times \frac{\Delta^{(i)}}{\left(X(t) - X^{(i)}\right)^2 + \left(\Delta^{(i)}\right)^2} \qquad (9)$$

The centres of the Lorentzians define the time at which water has been introduced / absorbed in the basin and the negative sign of the coefficients $A^{(i)}$ is interpreted as absorption. The widths $\Delta(i)$ describe time intervals of rainwater precipitation and / or diffusion. The results of the fitting are presented in the figures. The Fig.1 shows the decrease in the growth coefficient due to the drainage water, $Q_{BA1}(t)/V$, for BA1. In Fig.2 one can see that $Q_{BA2}(t)/V$, $Q_{BA3}(t)/V$ and $Q_{BA4}(t)/V$ are almost identical, and one can note that, except in the cases for zooplankton, $Q_{BZ}(t)/V$ and orthophosphate $Q_{PS}(t)/V$ (Figs. 3 and 4), which present large negative values, only positive values for the coefficients $A^{(i)}$ of the Lorentzians do exist. The Fig.5 shows the same, for the phosphorus in detritus, $Q_{PD}(t)/V$.

The advantage of the Lorentzian fit is to show at what time and for which time intervals the geological water dynamics for the drainage occurs. The negative terms of the expansion are interpreted as an important indication of absorption of these drainage elements by the lake.

Again, one points up the importance of separating the biochemical processes included in the non-linear functions $A_1^{(i)}(t)$, $A_2^{(i)}(t)$, $B_1^{(i)}(t)$, $B_2^{(i)}(t)$ and $B_3^{(i)}(t)$ of Eqs. (1) to (4) (including obviously the parameters) from the terms $Q_i$ and $Q_M$ which are fitted to the Lorentzians as previously described. This method thus shows, by the large negative values seen in Figs. 3 and 4, the relevance of time dependent absorption processes in the drainage, which describes the dynamics of the lake. Nevertheless, this time dependence has been disregarded in some existent approaches in the literature. Furthermore, these results suggest making geological experiments in the region of lakes in the North of Rio de Janeiro State to measure the significance of the time dependence associated with micrometeorological effects and the soil diffusion of water, in the dynamics of drainage water. These measurements could be performed using, for example, radioactive tracers.

**Acknowledgements**
The authors acknowledge J. Thadeu Cavalcante for his kind help by developing software for extracting the experimental and theoretical data plotted in figures of Hongping and Jianyi (2002) and to C. da Silva for a critical reading of the manuscript.

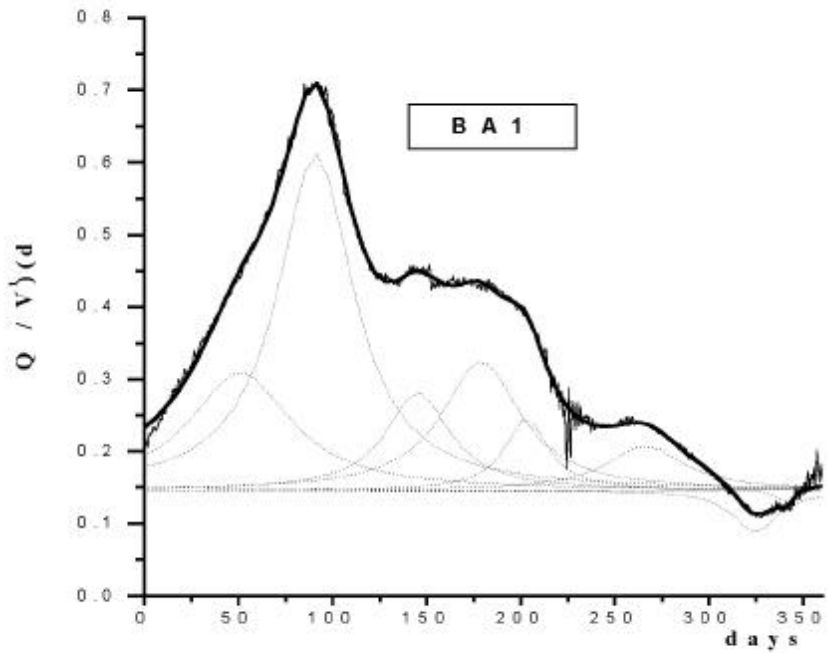

**Figure 1** – The irregular curve corresponds to the numerical solution for the decrease in the growth coefficient due to the drainage water of the algae Cyanophyta, $Q_{BA_1}(t)/V$ (see Eq. 1). The dashed lines are the eight Lorentzians distribution functions, which the sum gives the fit showed in the solid line.

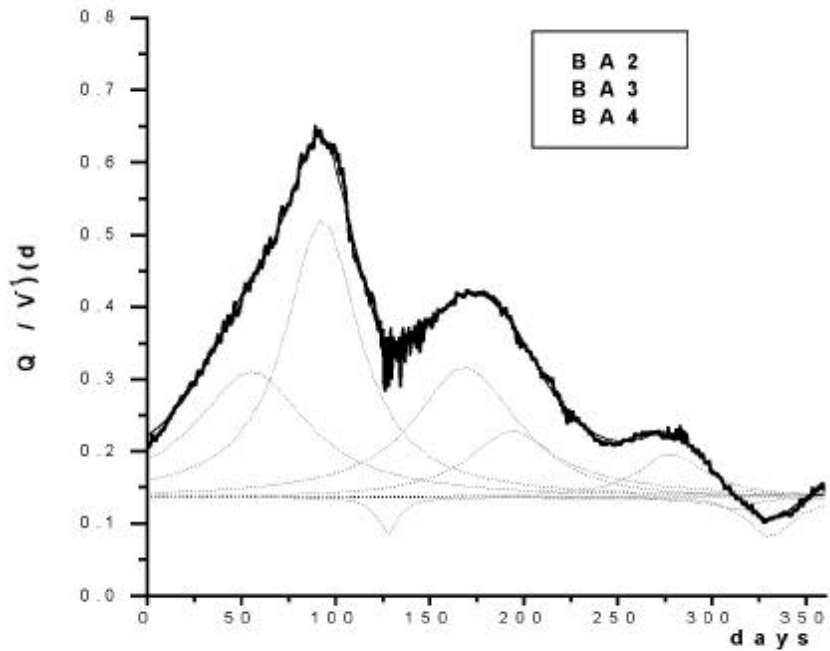

**Figure 2** - The irregular curve corresponds to the numerical solution for the decrease in the growth coefficient due to the drainage water of the algae Chlorophyta, Cryptophyta and Bacillariophyta, $Q_{BA_i}(t)/V$ (i = 2,3,4) respectively (see Eq. 1). (These three curves are almost indistinctive). Thus, it is shown only the eight Lorentzians distribution functions for $BA_2(t)$ (dashed lines), and the fit given by their sum (solid line).

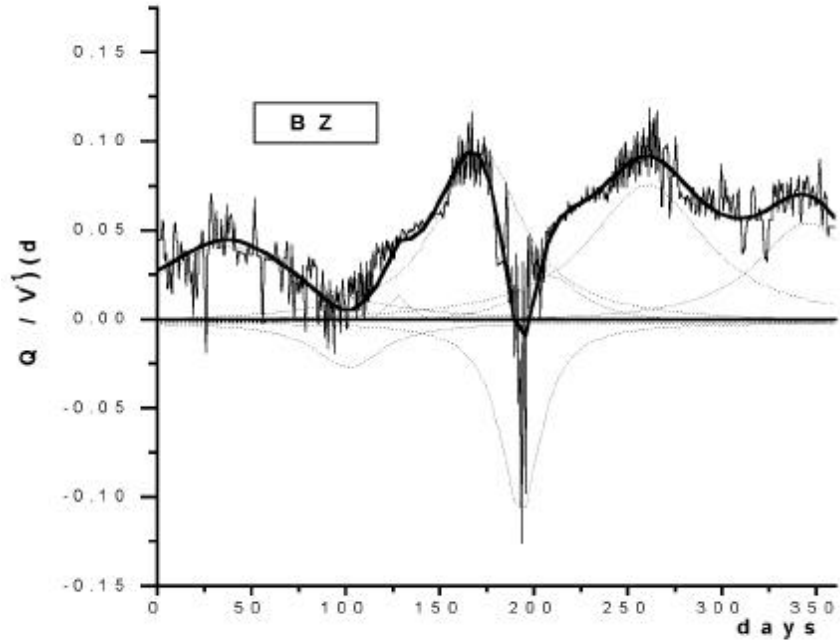

**Figure 3** - The irregular curve corresponds to the numerical solution for the decrease in the growth coefficient due to the drainage water of the zooplanktons, $Q_{BZ}(t)/V$ (see Eq. 2). The dashed lines are the eight Lorentzians distribution functions, which the sum gives the fit showed in the solid line.

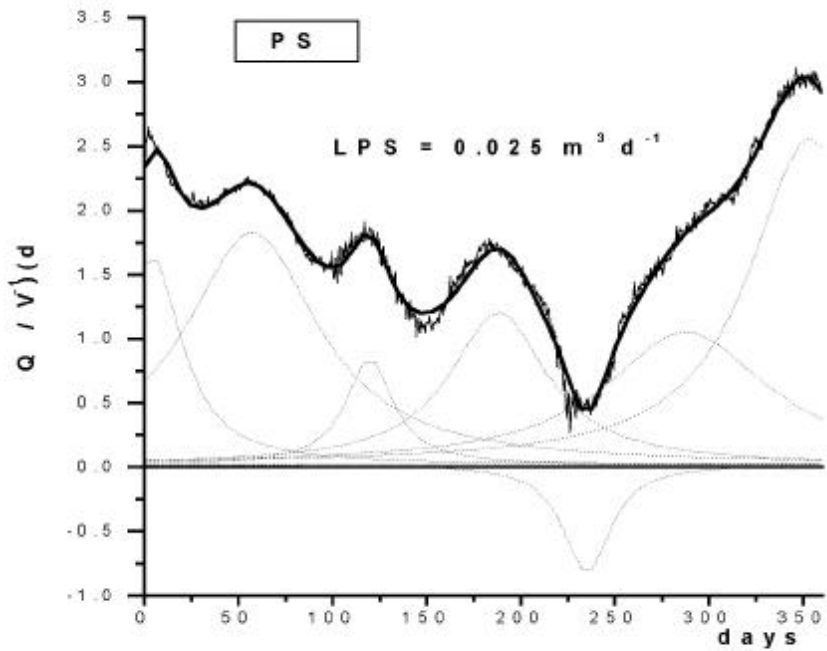

**Figure 4** - The irregular curve corresponds to the numerical solution for the decrease in the growth coefficient due to the drainage water of the ortophosphate, $Q_{PS}(t)/V$ (see Eq. 3). The dashed lines are the eight Lorentzians distribution functions, which the sum gives the fit showed in the solid line.

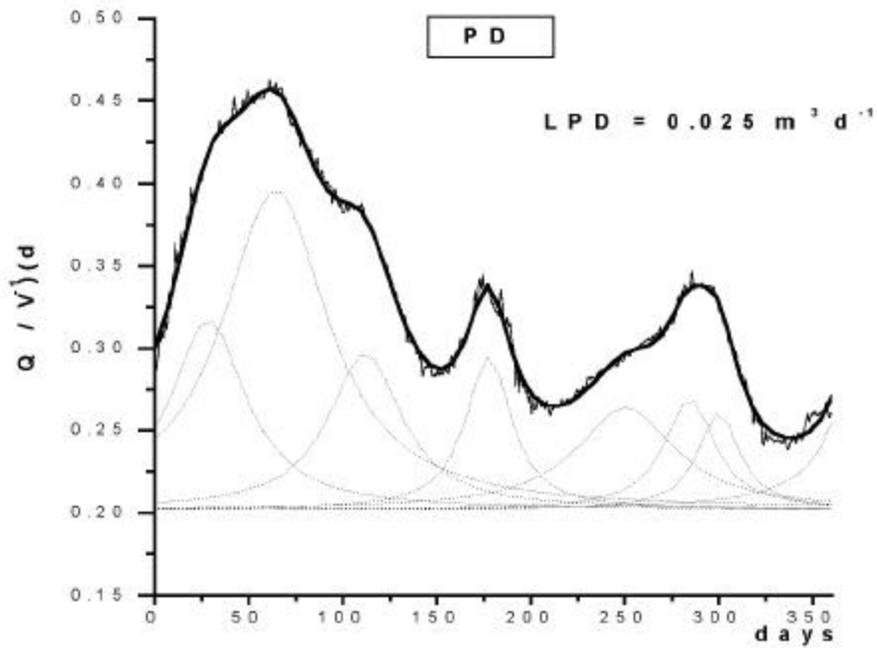

**Figure 5** - The irregular curve corresponds to the numerical solution for the decrease in the growth coefficient due to the drainage water of the phosphorus in detritus, $Q_{PD}(t)/V$ (see Eq. 4). The dashed lines are the eight Lorentzians distribution functions, which the sum gives the fit showed in the solid line.